\documentclass[twocolumn,showpacs,amsmath,amssymb, superscriptaddress]{revtex4}

\usepackage{amsmath}
\usepackage{amssymb}
\usepackage{amsfonts}

\usepackage{graphicx}
\usepackage{float}
\usepackage{color}

\usepackage{amsthm}

\DeclareMathOperator{\Tr}{Tr}

\DeclareMathOperator{\U}{\it U}
\DeclareMathOperator{\SU}{\it SU}
\DeclareMathOperator{\su}{\mathfrak s\mathfrak u}

\newcommand{\angl}[1]{{\left\langle #1 \right\rangle}}

\makeatletter
\def\loweq@align#1#2{\lower.6ex\vbox{\baselineskip\z@skip\lineskip\z@
    \ialign{$\m@th#1\hfil##\hfil$\crcr#2\crcr=\crcr}}}
\def\lowsim@align#1#2{\lower.6ex\vbox{\baselineskip\z@skip\lineskip\z@
    \ialign{$\m@th#1\hfil##\hfil$\crcr#2\crcr\sim\crcr}}}
\def\geqq{\mathrel{\mathpalette\loweq@align >}}
\def\leqq{\mathrel{\mathpalette\loweq@align <}}
\def\grsim{\mathrel{\mathpalette\lowsim@align >}}
\def\lesssim{\mathrel{\mathpalette\lowsim@align <}}
\def\gsim{\mathrel{\mathpalette\lowsim@align >}}
\def\lsim{\mathrel{\mathpalette\lowsim@align <}}
\makeatother
\newcommand{\grless} 
{ {\, \raise-.24em\hbox{$<$} \hspace{-0.8em} \raise.31em\hbox{$>$}\, } }
\newcommand{\lessgr} 
{ {\, \raise-.24em\hbox{$>$} \hspace{-0.8em} \raise.31em\hbox{$<$}\, } }
\newfont{\bg}{cmr10 scaled\magstep4}                    
\newcommand{\bigzerou}{\smash{\lower1.7ex\hbox{\bg 0}}}

\newcommand{\crl}[1]{[-\infty,\infty]}

\newcommand{\ket}[1]{|{#1}\rangle}

\newcommand{\Ref}[1]{(\ref{#1})}

\newcommand{\da}[1]{#1^\dag}

\newcommand{\alg}[1]{{\mathfrak #1}}

\newcommand{\av}[1]{\langle#1\rangle}

\begin{document}

\title{Time-optimal CNOT between indirectly coupled qubits in a linear Ising chain}
  \author{Alberto Carlini}
 \email{acarlini@mfn.unipmn.it}
 \affiliation{Dipartimento di Scienze e Tecnologie Avanzate, Universita' del Piemonte Orientale, Alessandria, Italy}
 \affiliation{Istituto Nazionale di Fisica Nucleare, Sezione di Torino, Gruppo Collegato di Alessandria, Italy}
\author{Akio Hosoya}
 \email{ahosoya@th.phys.titech.ac.jp}
 \affiliation{Department of Physics, Tokyo Institute of
 Technology, Tokyo, Japan}
\author{Tatsuhiko Koike}
 \email{koike@phys.keio.ac.jp}
 \affiliation{Department of Physics, Keio University, Yokohama, Japan}
\author{Yosuke Okudaira}
 \email{okudaira@th.phys.titech.ac.jp}
 \affiliation{Department of Physics, Tokyo Institute of
 Technology, Tokyo, Japan}

\date{September 29, 2010}
\begin{abstract}
We give analytical solutions for the time-optimal synthesis of entangling gates between indirectly coupled qubits 1 and 3 in a linear spin chain of three qubits subject to an Ising Hamiltonian interaction with a symmetric coupling $J$ plus a local magnetic field acting on the intermediate qubit. The energy available is fixed, but we relax the standard assumption of instantaneous unitary operations acting on single qubits. The time required for performing an entangling gate which is equivalent, modulo local unitary operations, to the $\mathrm{CNOT}(1, 3)$ between the indirectly coupled qubits 1 and 3 is $T=\sqrt{3/2}~J^{-1}$, i.e.
faster than a previous estimate based on a similar Hamiltonian and the assumption of local unitaries with zero time cost. Furthermore, performing a simple Walsh-Hadamard rotation in the Hlibert space of qubit 3 shows that the time-optimal synthesis of the $\mathrm{CNOT}^{\pm}(1, 3)$ (which acts as the identity when the control qubit 1 is in the state $\ket{0}$, while if the control qubit is in the state $\ket{1}$ the target qubit 3 is flipped as $\ket{\pm}\rightarrow \ket{\mp}$) also requires the same time $T$.

\end{abstract}  


\maketitle

\centerline{\sl To Alberto's mother, Giovanna Novaro Mascarello.}

\section{Introduction}

Quantum optimal control theory is by now a very well studied subject, both theoretically and experimentally, with increasing applications in the field of quantum computing and information (for an updated review see, e.g., \cite{brif} and references therein).
Quantum optimal control techniques aim, e.g., at finding either the best quantum state evolution or the best unitary operator evolution with respect to some fixed cost, which can be assumed to be a fidelity with respect to a target (a quantum state or a gate), the purity of the target state, etc.
Time-optimal quantum computation,  where the cost to be optimized is the time to achieve a given quantum evolution, has also recently become a hot topic in quantum control and quantum information theory
\cite{khaneja}-\cite{tanimura}.
For example, minimization of the physical time to achieve a given unitary transformation is relevant for the design of fast elementary gates and provides a more physical ground to describe the complexity of quantum algorithms than the standard concept of gate complexity, which gives the number of elementary gates used in a quantum circuit \cite{chuangnielsen}.
Also the literature in time-optimal quantum control is rapidly growing and it is not the purpose of this paper to give a complete review.
For instance, Refs. \cite{khaneja}-\cite{zhang} discuss the time 
optimal generation of unitary operations for a small number of qubits using Lie group methods, the theory of sub Riemannian geometry, the Pontryagin maximum principle 
and assuming that one-qubit operations can be performed arbitrarily fast.
The time-optimal synthesis of unitary transformations (quantum gates) between two coupled qubits has been thoroughly discussed \cite{kgb}-\cite{yuan3}.
The time-optimal evolution of quantum states for qubits and qutrits has been also investigated, e.g., in \cite{boscain}-\cite{boscain2}, that of a 2-level dissipative system, e.g., in \cite{sugny},
while the time-optimal generation of cluster states has been considered by \cite{fisher}.
The extension of the Lie algebraic methods to the coupling of slow and fast systems can be found in 
\cite{zeier}.
Earlier bounds on the time complexity of generating 2-qubit unitary gates can be found in \cite{childs},
while lower bounds on the time complexity of n-qubit gates are given in \cite{zeier2} and upper bounds
on the time complexity of certain n-qubit gates via several coupling topologies are numerically described in \cite{schulte}.
The relationship between time complexity and gate complexity has been also investigated in \cite{nielsen3}.
Nielsen et al. \cite{nielsen1}-\cite{gu} proposed a criterion for optimal quantum computation in terms of a certain geometry in Hamiltonian space, and showed that the quantum gate complexity is related to optimal control cost problems.
An adiabatic solution to the optimal control problem in holonomic quantum computation has also been found in \cite{tanimura}.
Finally, numerical methods for the design of optimal quantum control evolutions have been proposed,
based on the gradient ascent algorithm \cite{grape} and on the non linear Krotov algorithm \cite{dominy}.

Most of the above works share the assumption that one-qubit gates have zero time cost.
The present authors described a theoretical framework for time-optimal quantum computing based on the action principle where such an assumption is not necessary, and named it the quantum brachistochrone (from hereon abbreviated as QB \cite{greek}).
The variational principle is formulated for the time-optimal evolution of a quantum system whose Hamiltonian is subject to a set of constraints (e.g., a finite energy or magnetic field, certain qubit interactions are forbidden) and defines a boundary value problem with fixed initial 
and final quantum states (or unitary transformations). 
The QB has been studied for quantum state evolution in the case of pure \cite{pure} and mixed states
\cite{mixed}, and for the optimal realization of unitary transformations between the identity and a given target quantum gate \cite{unitary}.
The latter is particularly relevant for the standard quantum computation paradigm since a whole algorithm may be
reduced to a sequence of unitary transformations between intermediate states and a final measurement.
The more realistic situation where the target quantum state (gate) can be reached within a finite, tolerable error (a fidelity larger than a specified value) has also been addressed in \cite{fidelity}.
The quantum brachistochrone problem always reduces to solving a fundamental equation, which can be easily written down once the constraints for the Hamiltonian of the quantum system are known, with given initial boundary conditions, and an equation for the Lagrange multiplier which enforces the dynamical law for the quantum system (the Schr\"odinger equation for closed systems or, e.g., a master equation
for Markovian open systems), with given final boundary conditions. 
Two of us also studied numerically the time complexity of generating unitaries acting on $n$ qubits via a Hamiltonian which contains only one and two qubit interaction terms \cite{complexity}. 
Our research also triggered an amount of related works \cite{brodyhook}-\cite{nesterov}. 
For example, the authors of \cite{borras2}-\cite{zhao} considered the problem of the generation of multipartite entanglement during the QB evolution of quantum states and unitaries, while those of 
\cite{bbjm}-\cite{nesterov} studied the QB in the context of non Hermitian quantum mechanics
(for a review of the latter works see, e.g., \cite{benbrod}).

More recently, Khaneja et al.~\cite{khaneja2} considered the problem of the efficient synthesis of the controlled-{\scshape NOT} gate ($\mathrm{CNOT}(1, 3)$) between two qubits indirectly coupled via an Ising type coupling to a third qubit, and where the single qubits can be separately addressed via instantaneous local unitaries. 
Their implementation requires a time $T\simeq 1.253 J^{-1}\simeq \sqrt{\pi/2}~J^{-1}$, where $J$ is the Ising
coupling between the qubits (1, 2) and (2, 3) in a linear coupling topology, and it was related to the computation of a geodesic on the surface of a sphere with a special metric and to the synthesis of a particular entangling gate called $U^s_{13}$.
Subsequently, similar methods were used to extend these results to the case of unequal Ising couplings between the indirectly coupled qubits \cite{yuan} and to the case of a spin chain with $n$ qubits \cite{yuan2}.
The time-optimal synthesis of interactions between qubits which are indirectly coupled via an intermediate qubit is a typical scenario in quantum information processing (e.g., distant spins on spin chains \cite{kane}, bus-qubit electrons mediating interactions of two nuclear spins \cite{mehring} or NMR experiments correlating the frequencies of indirectly coupled qubits \cite{ernst}). The synthesis of such indirect couplings is usually time costly (e.g., it may require concatenation of two-qubit operations on the directly coupled qubits) and therefore it is also easily prone to decoherence effects and a degradation of the gate fidelity,  both critical effects in practical implementations of quantum computing.

In this work we further investigate the problem of the time-optimal generation of the same entangling quantum gates,  $U^s_{13}$  and $\mathrm{CNOT}(1, 3)$, between two indirectly coupled qubits in three-qubit systems with linear coupling topology and with similar Ising coupling Hamiltonian as that of \cite{khaneja2}.
We find that, if the available Hamiltonian is made of an interacting piece of the Ising type exactly as in the model of \cite{khaneja2} plus a local magnetic field acting on the intermediate qubit, and is subject to the constraint of a finite energy, then the $U^s_{13}$ gate can be optimally realized in a time $T=  \sqrt{3/2}~J^{-1}$, which is faster than the time required via the construction of \cite{khaneja2}. 
Furthermore, using a slightly modified interaction Hamiltonian (obtained via a simple change of basis for 
the Hilbert space of one of the indirectly coupled qubits, qubit 3) and the same energy constraint, we
find that the gate $\mathrm{CNOT}^{\pm}(1, 3)$ (which acts as the identity when the control qubit 1 is in the state $\ket{0}$, while if the control qubit is in the state $\ket{1}$ the target qubit 3 is flipped as $\ket{\pm}\rightarrow \ket{\mp}$) can be also optimally generated in the same time $T=  \sqrt{3/2}~J^{-1}$ as the entangling gate $U^s_{13}$ .
The paper is organized as follows.
In Section I we briefly review the main features of the QB formalism for the time-optimal synthesis 
of unitary quantum gates.
In Section II we discuss the problem of the efficient generation of the gate $U^s_{13}$ between the indirectly coupled boundary qubits of a three-linear qubit system subject to an Ising interaction and a local control
available for the intermediate qubit, when a finite energy is available.
Section III is devoted to the study of the time-efficient generation of the $\mathrm{CNOT}^{\pm}(1, 3)$ gate with the slightly modified interaction Hamiltonian plus the same local operation on qubit 2 and the same energy available.
Finally, Section IV is devoted to the summary and discussion of our results. 

\section{Quantum Brachistochrone}
We want to find the time-optimal way to generate a target unitary operation $U_f\in \U(N)$ (modulo physically irrelevant overall phases) by controlling an Hamiltonian $H(t)$ and evolving a unitary operator $U(t)$, both obeying the Schr\"odinger equation. 
We assume that $H$ is controllable within a certain available set, dictated either by theoretical conditions (e.g., only certain interactions among qubits are allowed) or by experimental requirements (e.g., a finite energy or a finite magnetic field).
At least the `magnitude' of the Hamiltonian must be bounded,  otherwise any gate $U_f$ might be realized in an arbitrarily short time simply by rescaling the Hamiltonian \cite{unitary}. 
Physically this corresponds to the fact that one can afford only a  finite energy in the experiment. 

The time-optimality problem is formulated using the action \cite{mixed,unitary}:
\begin{align}
  \label{eq-action}
  S(U,H; \alpha, \Lambda,\lambda_j) &:=\int_0^1 d\tau \left[ \alpha +L_S+L_C\right]
  \\
  \label{eq-LS}
  L_S &:=\av{\Lambda,i \tfrac{dU}{d\tau}\da U- \alpha H},
  \\
  \label{eq-LC}
  L_C &:= \alpha   \sum_j{\lambda_j}f^j(H), 
\end{align}
where $\angl{A,B}:=\Tr (\da AB)$ and the Hermitian operator
$\Lambda(\tau)$ and the real functions $\lambda_j(\tau)$ are Lagrange multipliers \cite{invariance}. 
The quantity $\alpha$ is the time cost, and it may be interpreted \cite{mixed} as a positive independent dynamical variable (a "lapse" function) which measures the physical time $t:=\int \alpha(\tau) d\tau$ lapsed in each infinitesimal interval $d\tau$ of the parameter time $\tau$.
Variation of $L_S$ by $\Lambda$ gives the Schr\"odinger equation:
\begin{align}
  \label{eq-Sch}
  i\frac{dU}{dt}=HU, \quad\text{or}\quad 
  U(t)={\mathcal  T}e^{-i\int^t_0 Hdt}, 
\end{align}
where $\mathcal T$ is the time ordered product. 
Variation of $L_C$ by $\lambda_j$ leads to the constraints for $H$:
\begin{align}
  f_j(H)=0. 
\label{constraints}  
\end{align}
In particular, the finite energy condition for a system of $\log N$ qubits can be written as: 
\begin{align}
f_0(H) :=\tfrac{1}{2}[\Tr(H^2)- N\omega^2]=0,
\label{eq-normH}
\end{align} 
where $\omega$ is a constant. 
Since overall phases are irrelevant in quantum mechanics, it is natural to
consider the time-optimal evolution of unitary operators belonging to the group $\U(N)/\U(1)\simeq \SU(N)$
(i.e., the dynamics is generated by a traceless Hermitian Hamiltonian). 

Then, we introduce the operator:
\begin{align}
  \label{eq-F-def}
  F:=\frac{\partial L_C}{\partial H}, 
\end{align}
and from \Ref{eq-normH} and \Ref{eq-F-def} we obtain 
\begin{align}
  \label{eq-F}
  F= \lambda_0 {H}+F'.
\end{align}  
If the other constraints are linear and homogeneous in $H$, 
i.e. $F'=\sum_j\lambda_jg_j$ with $g_j\in\su(N)$,
then we have \cite{unitary} $\Tr (HF')=0$
and it is easy to show \cite{mixed} that the Lagrange multiplier $\lambda_0$ in \Ref{eq-F}  is a constant.

From the variation of $S$ by $\alpha$, upon using \Ref{eq-normH} and \Ref{eq-F}, one gets $1=\Tr (HF)=\lambda_0 \Tr(H^2)=\lambda_0 N\omega^2$, which determines the constant $\lambda_0$. 

Finally, variation of $S$ by $U$ and some elementary algebra give the 
{\em quantum brachistochrone equation} 
\begin{align}
  \label{eq-fund}
  i\frac{dF}{dt}= [H, F],
\end{align}
The Òquantum brachistochroneÓ together with the constraints define a boundary-value problem for the evolution of the unitary operator $U(t)$ with fixed initial ($U(t=0)=1$, where $1$ is the identity matrix) and final conditions $(U(t=T)=U_f$, where $T$ is the optimal time duration necessary to achieve the target gate $U_f$).
The quantum brachistochrone is a set of first-order (non linear) differential equations which can always be solved in principle, e.g. numerically, and it is universal, as it holds also in the case of time-optimal evolution of pure \cite{pure} and mixed \cite{mixed} quantum states.

More in details, for a given target gate $U_f$, the procedure to find the optimal Hamiltonian $H$ and the optimal time duration $T$ consists of the following stages: (i) specify the constraint functions $f_j(H)$ for the available Hamiltonian; (ii) write down and solve the quantum brachistochrone \Ref{eq-fund} together with the constraints \Ref{constraints} to obtain $H_{OPT}(t)$; (iii) integrate the Schr\"odinger equation \Ref{eq-Sch} with $U(0)=1$ to get $U_{OPT}(t)$; (iv) fix the integration constants in $H_{OPT}(t)$ by imposing the condition that $U_{OPT}(T)$ equals $U_f$ modulo a global (physically irrelevant) phase, i.e., 
 \begin{align}
  U_{OPT}(T)=e^{i\chi}~U_f, 
  \label{targetU}
\end{align}
where $\chi$ is some real number.

\section{Ising Hamiltonian and Time-Optimal Entangler $U^s_{13}$}

Now we apply the general QB formalism summarized in the previous Section to the case of 
a physical system of three qubits represented by three spins (labeled by  a
superscript $a\in \{1, 2,  3\}$) interacting via an Ising Hamiltonian with time independent couplings 
$J_{12}, J_{23}$ and subject  to a local and controllable magnetic field $B^2_i(t)$ ($i=x, y, z$). 
In other words, we choose the three-qubit Ising Hamiltonian,
\begin{eqnarray}
 H(t)\!\!& :=&\!\! \frac{\pi}{2}[J_{12}(t)\sigma_z^{1}\sigma_z^{2} +J_{23}(t)\sigma_z^{2}\sigma_z^{3}] 
 +\vec{B}(t)\cdot \vec{\sigma}^{2},
\label{ising}
\end{eqnarray}
where we have used the simplified notation, e.g., $\sigma_i^1\sigma_j^2:=\sigma_i\otimes \sigma_j\otimes 1$,
$\sigma_i^2\sigma_j^3:=1\otimes \sigma_i\otimes \sigma_j$,
$\sigma_i^{2} :=1\otimes \sigma_i\otimes 1$ and $\sigma_i$ are the Pauli
operators \cite{chuangnielsen}.
We further assume that the Ising couplings in \Ref{ising} are equal and (a positive) constant, i.e. $J_{12}(t)=J_{23}(t):=J>0$ \cite{otherway}.
This is formally enforced via the Lagrange multipliers $\nu_{zz}(t)$ and $\rho_{zz}(t)$ and, respectively, the 
associated constraints
\begin{eqnarray}
f_1&:=&\mathrm{Tr} (H\sigma_z^1\sigma_z^2)-4\pi J=0\label{cf1}\\
f_2&:=&\mathrm{Tr} (H\sigma_z^2\sigma_z^3)-4\pi J=0\label{cf2}.
\end{eqnarray}
Moreover, the finite energy condition \Ref{eq-normH} reads
\begin{equation}
 \vec{B}^2=\omega^2 -\frac{(\pi J)^2}{2}=\mathrm{const}.
 \label{const-H}
\end{equation}
We note that the interaction part of the Hamiltonian \Ref{ising} is exactly the same as $H_c$ in eq. (2) of \cite{khaneja2}, and that the local term in \Ref{ising} corresponds to the terms $H_A$ and $H_B$ of
\cite{khaneja2}.
Finally, the form \Ref{ising} of the physical Hamiltonian is guaranteed by the operator
\begin{eqnarray}
 F'(t)&=&\sum_{i, j, k}\lambda_{ijk}(t) \sigma_i^1 \sigma_j^2\sigma_k^3
 \nonumber \\
 & +& \sum_{i, j}[ \mu_{ij}(t) \sigma_i^1\sigma_j^3
  +  \nu_{ij}(t) \sigma_i^1\sigma_j^2
  + \rho_{ij}(t) \sigma_i^2\sigma_j^3]\nonumber\\
&+& \sum_i [\eta_i(t) \sigma_i^1 + \xi_i(t) \sigma_i^3] 
\label{fheisenberg}
\end{eqnarray}
where $\lambda_{ijk}(t), \mu_{ij}(t), \nu_{ij}(t), \rho_{ij}(t), \eta_i(t)$ and  $\xi_i(t)$ are Lagrange multipliers, and the indices
$\{i, j\}\in \{x, y, z\}$.

Our task is then to solve the quantum brachistochrone equation
\Ref{eq-fund}. 
Comparing the coefficients of
the generators of ${\alg su}(8)$ on both sides of \Ref{eq-fund}, with $F$ given by \Ref{eq-F}, $H$ given by \Ref{ising} and $F'$ given by \Ref{fheisenberg},
we find that the relevant quantum brachistochrone equations are:
\begin{eqnarray}
\dot{B}_x&=&-\pi J(\nu_{zy} +\rho_{yz})\label{f1}\nonumber\\
\dot{B}_y&=&\pi J(\nu_{zx} +\rho_{xz})\label{f2}\nonumber\\
\dot{B}_z&=&0\label{f3}\nonumber\\
\dot{\nu}_{zx}&=&-[\pi J\lambda_{zyz} +2(B_z\nu_{zy} -B_y\nu_{zz})]\label{f6}\nonumber\\
\dot{\nu}_{zy}&=&\pi J\lambda_{zxz} +2(B_z\nu_{zx} -B_x\nu_{zz})\label{f7}\nonumber\\
\dot{\nu}_{zz}&=&2(B_x\nu_{zy} -B_y\nu_{zx})\label{f4}\nonumber\\
\dot{\rho}_{xz}&=&-[\pi J\lambda_{zyz} +2(B_z\rho_{yz} -B_y\rho_{zz})]\label{f8}\nonumber\\
\dot{\rho}_{yz}&=&\pi J\lambda_{zxz} +2(B_z\rho_{xz} -B_x\rho_{zz})\label{f9}\nonumber\\
\dot{\rho}_{zz}&=&2(B_x\rho_{yz} -B_y\rho_{xz})\label{f5}\nonumber\\
\dot{\lambda}_{zxz}&=&-\pi J(\nu_{zy}+\rho_{yz}) +2(B_y\lambda_{zzz} -B_z\lambda_{zyz})\label{f10}\nonumber\\
\dot{\lambda}_{zyz}&=&\pi J(\nu_{zx}+\rho_{xz}) -2(B_x\lambda_{zzz} -B_z\lambda_{zxz})\label{f11}\nonumber\\
\dot{\lambda}_{zzz}&=&2(B_x\lambda_{zyz} -B_y\lambda_{zxz}\label{f12})
\label{feq}
\end{eqnarray}
As recalled in the previous Section, the first step in addressing the QB problem is to solve the fundamental equation \Ref{eq-fund} for the time-optimal Hamiltonian $H_{OPT}(t)$.
In other words, we need to solve eqs. \Ref{feq} together with the constraints \Ref{cf1}-\Ref{const-H} and with fixed parameters $J, \omega$. 
From the integral of the motion $B_z=\mathrm{const}$, eqs. \Ref{feq} and the energy constraint \Ref{const-H} we immediately obtain that $B_x^2+B_y^2 := B_0^2=\mathrm{const}$ while, from eqs. \Ref{feq}, we further find the integrals of the motion $\nu_{zz}+\rho_{zz}=\mathrm{const}$, $\lambda_{zzz}=\mathrm{const}$.
Exploiting these integrals of the motion and after some lengthy but elementary algebra, we find that the general and non trivial solution of \Ref{feq} is given by the Hamiltonian \Ref{ising} with the time-optimal magnetic field
\begin{align}
  \vec{B}_{OPT}(t)= \left ( \begin{array}{c}
 {B_0}\cos \theta(t)\\
 {B_0}\sin \theta(t)\\
   B_z   
  \end{array}\right )
  \label{bopt}
\end{align}
precessing around the $z$-axis with the frequency $\Omega$, where $\theta(t):=\Omega t +\theta(0)$ and $\Omega$ and $\theta(0)$ are integration constants.

The next step is then to integrate the Schr\"odinger eq. \Ref{eq-Sch} for $U_{OPT}(t)$, given that $H_{OPT}(t)$ is expressed by 
eqs. \Ref{ising} and \Ref{bopt}.
For this purpose, we exploit the following well known property for the rotation of the Pauli matrices:
\begin{align}
e^{-i\frac{\theta(t)}{2}\sigma_z}\sigma_x e^{i\frac{\theta(t)}{2}\sigma_z}=\cos \theta(t) \sigma_x +\sin \theta(t)\sigma_y,
\end{align}
and we rewrite the time-optimal Hamiltonian as
\begin{align}
H_{\mathrm{OPT}}(t)=e^{-i\frac{\theta(t)}{2}\sigma_z^2} H_0~ e^{i\frac{\theta(t)}{2}\sigma_z^2},
\label{hopt}
\end{align}
where we have introduced the constant operator 
\begin{eqnarray}
H_0:={B_0}\sigma_x^2 +\left [\frac{\pi J}{2}(\sigma_z^1 +\sigma_z^3)+B_z\right ]\sigma_z^2.
\end{eqnarray}
Furthermore, defining the transformed unitary operator
\begin{align}
\tilde{U}(t):=e^{i\frac{\theta(t)}{2}\sigma_z^2}U(t),
\label{tildeU}
\end{align}
we easily check that, since $U(t)$ should obey the Schr\"odinger equation \Ref{eq-Sch}, $\tilde{U}(t)$ should also satisfy
\begin{align}
i\frac{d\tilde{U}}{dt}=\tilde{H}\tilde{U}
\label{schrnew}
\end{align}
with the time-independent Hamiltonian
\begin{align}
\tilde{H}:=H_0-\frac{\Omega}{2}\sigma_z^2=\mathrm{const}.
\end{align}
We note that the constant Hamiltonian $\tilde{H}$ is diagonal in the 1,3 qubit subspace, i.e.
\begin{align}
\tilde{H}={B_0}\sigma_x^2 + B_D^{13} \sigma_z^2=\mathrm{const},
\label{tildeH}
\end{align}
where we have introduced the operator (in the 1,3 qubit subspace)
\begin{align}
B_D^{13}:=-\frac{1}{2}(\Omega -2B_z) 1 + \pi J ~\mathrm{Diag}[1, 0, 0, -1 ].
\label{bd13}
\end{align}

Then, solving eq. \Ref{schrnew} together with eq. \Ref{tildeH} for $\tilde{U}(t)$ and finally inverting \Ref{tildeU}, it is easy to check that the time-optimal unitary operator $U_{OPT}(t)$ evolves as 
\begin{align}
U_{\mathrm{OPT}}(t)=e^{-i\frac{\theta(t)}{2}\sigma_z^2}e^{-i\tilde{H}t}e^{i\frac{\theta(0)}{2}\sigma_z^2}.
\label{uopt}
\end{align}
In particular, the exponential of the constant Hamiltonian appearing on the right hand side of eq. \Ref{uopt}
is also diagonalized (in the 1, 3 qubit subspace) and can be expanded as
\begin{align}
e^{-i\tilde{H}t}=C_D^{13}(t)-iS_D^{13}(t)\tilde{H},
\label{exptildeH}
\end{align}
where we have introduced the following operators (acting in the 1,3 qubit subspace)
\begin{eqnarray}
S_D^{13}(t)&:=&\mathrm{Diag}[s_+(t), s_0(t), s_0(t), s_-(t),]\label{scoperators1}\\
C_D^{13}(t)&:=&\mathrm{Diag}[c_+(t), c_0(t), c_0(t), c_-(t)],
\label{scoperators2}
\end{eqnarray}
which depend upon the functions 
\begin{eqnarray}
s_{\pm}(t)&:=&\frac{\sin \omega_{\pm} t}{\omega_{\pm}}; ~~s_{0}(t):=\frac{\sin \omega_{0} t}{\omega_{0}},
\label{scfunctions1}\\
c_{\pm}(t)&:=&\cos \omega_{\pm} t;~~c_{0}(t):=\cos \omega_{0} t,
\label{scfunctions2}
\end{eqnarray}
and the constants
\begin{eqnarray}
\omega_{\pm}^2&:=&B_0^2+\frac{1}{4}[\Omega -2(B_z \pm \pi J)]^2,\label{omegafunctions1}\\
\omega_0^2&:=&B_0^2+\frac{1}{4}(\Omega - 2B_z)^2.
\label{omegafunctions2}
\end{eqnarray}
Then, inserting eq. \Ref{exptildeH} into eq. \Ref{uopt}, one obtains the following more explicit expression
for the time-optimal evolution of the unitary operator, i.e.
\begin{eqnarray}
U_{\mathrm{OPT}}(t)&=&\cos\frac{\Omega t}{2} ~C_D^{13}(t)-\sin\frac{\Omega t}{2}~B_D^{13}S_D^{13}(t)\nonumber\\
&-&i \biggl \{{B_0}S_D^{13}(t) [ \cos\phi(t) \sigma_x^2 
-\sin\phi(t)\sigma_y^2  ] \nonumber \\
&+& \biggl [\sin\frac{\Omega t}{2} C_D^{13}(t)+ \cos\frac{\Omega t}{2} B_D^{13}S_D^{13}(t)\biggr ]\sigma_z^2\biggr \},
\label{uoptexpl}
\end{eqnarray}
where $\phi(t)\equiv -[\theta(t)+\theta(0) ]/2$.

The expression \Ref{uoptexpl} for $U_{OPT}(t)$ still depends upon the integration constants $B_0, B_z$, $\Omega$ and $\theta(0)$ and the coupling $J$.
These constants, together with the optimal duration time $T$ of the evolution and the irrelevant global phase $\chi$, can be finally fixed by imposing the target condition \Ref{targetU}.

We are interested here in the time-optimal realization of the symmetric entangler gate $U^s_{13}$  (see eq. (4) in \cite{khaneja2}), i.e.
\begin{align}
U_f =U^s_{13}: = e^{-i\frac{\pi}{4}(\sigma^1_z\sigma^3_z+\sigma^1_z+\sigma^3_z)}.
\label{ufus}
\end{align}
This is diagonal in the 1, 3 qubit subspace and explicitly reads
\begin{align}
U_f =e^{i\frac{\pi}{4}}U_D^{13}, ~~~~U_D^{13}:= \mathrm{Diag}(-1, 1, 1, 1). 
\label{ufus1}
\end{align}

Imposing the target condition eq. \Ref{targetU} with $U_{OPT}(t)$ given by eq. \Ref{uoptexpl} and $U_f$ given by eqs. \Ref{uf1}-\Ref{uf2}, and separately equating the terms multiplying the Pauli operators $\sigma^2_x, \sigma^2_y$ and $\sigma^2_z$ and the identity operator $1^2$, respectively, we obtain the following set of conditions for operators acting on the 1,3 qubit subspace:
\begin{eqnarray}
0&=&B_0\cos\phi(T) S_D^{13}(T), \label{tar1}\\
0&=&B_0\sin\phi(T) S_D^{13}(T), \label{tar2}\\
0&=&\sin\frac{\Omega T}{2} C_D^{13}(T) + \cos\frac{\Omega T}{2} B_D^{13}S_D^{13}(T), \label{tar3}\\
e^{i\left [\chi +\frac{\pi}{4}\right ]}U^{13}_D&=&\cos\frac{\Omega T}{2} C_D^{13}(T)-\sin\frac{\Omega T}{2}B_D^{13}S_D^{13}(T).
\label{tar4}
\end{eqnarray}
From eq. \Ref{tar4} (imposing the reality of its left hand side term) we immediately obtain the value of the global phase $\chi =(k-1/4)\pi$, where $k\in \mathbb{Z}$.
Then, the only non trivial solution \cite{b0} for eqs. \Ref{tar1}-\Ref{tar2} is easily seen to be $S_D^{13}(T)=0$, which is possible, using eq. \Ref{scoperators1},  if and only if $s_+(T)=s_-(T)=s_0(T)=0$ and, upon comparing with eq. \Ref{scfunctions1}-\Ref{scfunctions2}, provided that
\begin{eqnarray}
\omega_{\pm}T= \pi n_{\pm}, ~~~~\omega_0T=\pi n_0,
\label{omegasolution}
\end{eqnarray}
where $n_{\pm}$ and $n_0$ are positive integers.
Substituting the solution \Ref{omegasolution} into eqs. \Ref{tar3} we also obtain 
the time-optimal integral of the motion $\Omega T=2\pi m$, where $m\in \mathbb{Z}$.
Finally, inserting the time-optimal formulas for $\chi$, $\Omega$ and $\omega_{\pm}, \omega_0$ into eq. \Ref{tar4}, we obtain the conditions
\begin{eqnarray}
n_+ = k_0 +2p; ~~n_ -= n_+ +2r-1; ~~n_0 = n_+ +2q-1
\label{npar}
\end{eqnarray}
with $k_0:=k+m+1$, $~p, q, r \in \mathbb{Z}$, and the parity of $n_0$ and $n_-$ the same and the opposite of that of $n_+$.

At this point we note that the formulas \Ref{omegafunctions1}-\Ref{omegafunctions2} can be easily inverted (using \Ref{omegasolution}) to write down the parameters $B_0, B_z$ and $J$ and the optimal time duration $T$ as functions of the integers  $n_{\pm}$ and $n_0$.
Introducing the following (odd-integer-valued) functions of the integers $n_{\pm}$ and $n_0$
in order to simplify the notation,
\begin{eqnarray}
f_{\pm}(n_+,  n_-,  n_0)&:=&n_+^2+n_-^2\pm 2n_0^2,\label{fpm}\\
f_0(n_+,  n_-)&:=&n_+^2-n_-^2,
\label{fpm0}
\end{eqnarray}
from eqs. \Ref{omegafunctions1}-\Ref{omegafunctions2} we thus obtain
\begin{eqnarray}
(JT)^2&=&\frac{f_-}{2}, \label{jt}\\
(B_0T)^2&=&\pi^2\left [n_0^2 - \frac{1}{8}\left(\frac{f_0}{\sqrt{f_-}}\right )^2\right ],\label{b0t}\\
(B_zT)^2&=&\pi^2\left (m - \frac{\sqrt{2}}{4}\frac{f_0}{\sqrt{f_-}}\right )^2.\label{bzt}
\end{eqnarray}
We immediately see that one can minimize $JT$ from eq. \Ref{jt} by minimizing $f_-$ as a function of the integers $k, p, q$ and $r$ (via eqs. \Ref{npar} and \Ref{fpm}). 
In other words, we can determine the optimal time duration $T$ of the quantum evolution necessary to realize the target gate $U^s_{13}$ as measured in terms of the coupling $J$. 
Then, substituting the time-optimal values of $k, p, q$ and $r$ into eqs. \Ref{fpm}-\Ref{fpm0} and 
subsequently into \Ref{b0t}-\Ref{bzt}, we obtain the time-optimal values of the integrals of the motion
$B_0$ and $B_z$.
Finally, we still have to impose the energy constraint \Ref{const-H} which, in terms of the functions $f_{\pm}$ and $f_0$, explicitly reads
\begin{eqnarray}
(\omega T)^2=\frac{\pi^2}{4}\left [f_+  - \frac{1}{2}\left [\frac{f_0}{\sqrt{f_-}}\right ]^2
+4\left [m - \frac{\sqrt{2}}{4}\frac{f_0}{\sqrt{f_-}}\right ]^2 \right ].
\label{econd}
\end{eqnarray}
Substituting for the time-optimal values of $k, p, q$ and $r$ into eqs. \Ref{fpm}-\Ref{fpm0} and 
\Ref{econd}, this constrains the values of the coupling $J$ as a function of the (given) energy parameter $\omega$.

More explicitly, let us look first for the minima of $T$ via eq. \Ref{jt}.
For a start, we note that, due to the parity properties of $n_+, n_-$ and $n_0$ (eqs. \Ref{npar}), the function $f_-(n_+, n_-, n_0)$ is always an odd integer. 
Moreover, its minimum value $f_{-\mathrm{MIN}}=1$ in principle may be achieved in two cases, i.e. either when $n_+$ is even ($k_0$ even) or odd ($k_0$ odd).
It is immediate to check that the case of $n_+$ even (and therefore, via \Ref{npar}, $n_-$ and $n_0$ odd) is not possible \cite{us13}. 
Furthermore, in the case in which $n_+$ is odd, defining $k_0:=2k_O+1$ (with $k_O\in \mathbb{Z}$), the condition $f_{-\mathrm{MIN}}=1$ translates into $(p+k_O)(2r-4q+1)=2q^2-r^2$.
Substituting  $f_{-\mathrm{MIN}}=1$ into eq. \Ref{b0t} we can rewrite $(B_0T)^2=-16\pi^2\{[q+r+2(p+k_O)]^2-1/8\}[(q-r)^2-1/8]$.
Imposing the reality condition, i.e. that $(B_0T)^2\geq 0$, after a simple algebra  
we finally obtain that the only possible solution for $f_{-\mathrm{MIN}}=1$ would be given by 
$(1-2q)[1+2q+4(p+k_O)]=1$, with $r=q$ or $r=-q/(1-2q)$ and $q\not = 0, 1$, which no set of integers $\{k_O, p, q, r\}$ 
can ever satisfy.
The next step is to check whether the next-to minimum value of  $f_{-\mathrm{NM}}=3$ may be consistent with the reality condition for $(B_0T)^2\geq 0$.
Again, it is immediate to verify that the case of  $n_+$ odd is impossible \cite{foot1}.
However, in the case of $n_+$ even, by defining $k_0:= 2k_E$  (with $k_E\in \mathbb{Z}$), the condition $f_{-\mathrm{NM}}=3$ translates into $(p+k_E)(2r-4q+1)=1-r(r-1)+2q(q-1)$.
Calculating \Ref{b0t} for $f_{-\mathrm{NM}}=3$ and $n_+$ even, we obtain $(B_0T)^2=-(\pi^2/3)\{[4(p+k_E)+2q-1]^2-3/2\}
[(2q-1)^2-3/2]$.
The reality condition $(B_0T)^2\geq 0$ now can be satisfied either for the values of the integers $p+k_E=1$, with $q=0, -1$ and $r=0, -1$, or for $p+k_E=-1$, with $q=1, 2$ and $r=1, 2$.
Summarizing, we have found that the optimal duration time required to realize a $U^s_{13}$ gate with the Ising Hamiltonian \Ref{ising} is given by 
\begin{align}
T=\sqrt{\frac{3}{2}}~J^{-1},
\label{topt}
\end{align}
the corresponding time-optimal magnetic fields are 
\begin{eqnarray}
|B_0|&:=&\frac{\sqrt{5}}{2\sqrt{3}}~\pi J,\label{b0z1}\\
|B_z|&:=&\sqrt{\frac{2}{3}}~\pi J\left | m-\sqrt{\frac{3}{8}}\right |,
\label{b0z2}
\end{eqnarray}
while the time-optimal precession frequency of the magnetic field is given by 
\begin{eqnarray}
|\Omega|= 2\sqrt{\frac{2}{3}}~\pi m~J,
\label{Omegaopt}
\end{eqnarray}
where the integer $m$ and the phase $\theta(0)$ are still arbitrary.
Moreover, the coupling $J$ and the energy parameter $\omega$ are constrained (from eq. \Ref{econd})
by the following condition 
\begin{eqnarray}
\frac{J}{|\omega |} =\frac{2\sqrt{3}}{\pi}\left [11+8\left (m-\sqrt{\frac{3}{8}}\right )^2\right ]^{-1/2}.
\label{jomega}
\end{eqnarray}


\section{Modified Hamiltonian and Time-Optimal $\mathrm{CNOT}(1, 3)$}

An analysis similar to that performed in the last Section can be done in the following two situations.
On the one hand, one can assume to be in the physical situation in which the Hamiltonian \Ref{ising} is available to a given experimentalist $O$ working with the standard computational basis, which in our three-qubit model corresponds to the Hilbert space spanned by the states $\{\ket{0}, \ket{1}\}\otimes \{\ket{0}, \ket{1}\}\otimes \{\ket{0}, \ket{1}\}$.
Then, we can think of another experimentalist $O'$ who performs measurements in the basis which corresponds the Hilbert space spanned by the states $\{\ket{0}, \ket{1}\}\otimes \{\ket{0}, \ket{1}\}\otimes\{\ket{+}, \ket{-}\}$, i.e. the basis where the rotated states $\{\ket{+}, \ket{-}\}$ (with $\ket{+}:=W\ket{0}=(\ket{0}+\ket{1})/\sqrt{2}; ~\ket{-}:=W\ket{1}=(\ket{0}-\ket{1})/\sqrt{2}$ and $W$ is the Walsh-Hadamard transform) are used for qubit 3, while qubits 1 and 2 use the standard computational basis.
More formally, this situation is equivalent to the second experimentalist seeing an effective, rotated Hamiltonian given by
 \begin{align}
 H^\prime(t)\!\!& :=&\!\! \frac{\pi}{2}[J_{12}(t)\sigma_z^{1}\sigma_z^{2} +J_{23}(t)\sigma_z^{2}\sigma_x^{3}] +\vec{B}(t)\cdot \vec{\sigma}^{2}.
\label{ising2}
\end{align}
On the other hand, one may think of the situation in which one experimentalist who can perform measurements in the standard computational basis for the three-qubit system, but 
with the Hamiltonian \Ref{ising2} available.

In both cases, the Hamiltonians \Ref{ising} and \Ref{ising2} are related by the transformation
\begin{align}
H^\prime(t)=VH(t)V,
\label{rotatedh}
\end{align}
where we have introduced the operator $V:=1\otimes 1\otimes W$.
Then, one may formulate a QB problem using an action principle similar
to that of the previous Section.
In other words, one obtains the quantum brachistochrone eq. \Ref{eq-fund}, and the
form of the available Hamiltonian \Ref{ising2} is guaranteed by the operator $F^\prime$ formally given by eq. \Ref{fheisenberg} with the same set of Lagrange multipliers,
but with the constraint \Ref{cf2} now replaced by 
\begin{eqnarray}
f^\prime_2&:=&\mathrm{Tr} (H\sigma_z^2\sigma_x^3)-4\pi J=0\label{cf22}
\end{eqnarray}
(while the energy constraint \Ref{const-H} is the same).
A lengthy but simple algebra shows that the quantum brachistochrone equations are 
again given by \Ref{feq}, with the exception that now we have to replace everywhere
\begin{align}
\rho_{iz}\rightarrow \rho_{ix}~;~\lambda_{ziz}\rightarrow \lambda_{zix}~;~\forall i\in \{x, y, z\}.
\end{align}
Then, we can follow the same procedure as in the previous Section and find that eqs. 
\Ref{feq} admit the integral of the motion $B_x^2+B_y^2=(B_0^\prime)^2$ and the general solution \Ref{bopt}.
In particular, the time-optimal Hamiltonian can be seen to become 
\begin{align}
H^\prime_{OPT}(t)=VH_{OPT}(t)V,
\end{align}
where $H_{OPT}(t)$ is given by eq. \Ref{hopt}.
The corresponding time-optimal evolution operator is then given by $U'_{OPT}(t)=VU_{OPT}(t)V$,
with the $U_{OPT}(t)$ of eq. \Ref{uoptexpl}, and with the operators $B_D^{13}, S^{13}_D$ and $C^{13}_D$ always given by eqs. \Ref{bd13} and \Ref{scoperators1}-\Ref{scoperators2}, respectively.

Now the goal is to time-optimally synthesize the gate $\mathrm{CNOT}(1, 3)$, i.e.
\begin{align}
U_f:=  CNOT(1, 3) = e^{-i\frac{\pi}{4}(1+ \sigma^1_z\sigma^3_x-\sigma^1_z-\sigma^3_x)},
\label{uf1}
\end{align}
which can be diagonalized (in the 1, 3 qubit subspace) as
\begin{align}
U_f=V{U'}_D^{13} V,
\label{uf2}
\end{align}
where  we have introduced the operator (acting in the 1,3 qubit subspace)
${U'}_D^{13}:= \mathrm{Diag}(1, 1, 1, -1) $.
Following the same methods of the previous Section, we then find that the target condition \Ref{targetU} is equivalent to impose again the operator conditions \Ref{tar1}-\Ref{tar4}, with the only difference that in \Ref{tar4} we have to replace
$\exp (i\chi+ \pi/4)U^{13}_D\rightarrow \exp(i\chi^\prime){U^\prime}^{13}_D$.
This implies that the optimal value of the global phase is given now by
$\chi^\prime =k^\prime \pi$, with $k^\prime\in \mathbb{Z}$, while the optimal
$\Omega$ is the same as in the previous Section.
The non trivial solution for eqs. \Ref{tar1}-\Ref{tar2} is again given by $s_+(T)=s_-(T)=s_0(T)=0$, and consequently, by eq. \Ref{omegasolution} for the same functions $\omega_{\pm}$ and $\omega_0$ given by eqs. \Ref{omegafunctions1}-\Ref{omegafunctions2}.
However, substitution of the optimal values of $\chi$ and $\Omega$ and $\omega_{\pm}$ and
$\omega_0$ into eq. \Ref{tar4} now gives, instead of eqs. \Ref{npar},
\begin{eqnarray}
n^\prime_+ =k^\prime_0 +2p^\prime ;~~n^\prime_ -= n^\prime_+ +2r^\prime +1; ~~
n^\prime_0 = n^\prime_+ +2q^\prime ,
\label{npar2}
\end{eqnarray}
with $k^\prime_0:=k^\prime+m$, $~p^\prime, q^\prime, r^\prime \in \mathbb{Z}$, and the parity of $n^\prime_0$ and $n^\prime_+$ the same and the opposite of that of $n^\prime_-$.
Considerations similar to those made in the previous Section lead to the expressions
\Ref{jt}-\Ref{econd}, and to the minimization of the evolution time $T$ at $f_{-\mathrm{NM}}=3$ (with $f_\pm$ and $f_0$ always defined by \Ref{fpm}-\Ref{fpm0}).
In this case the condition $f_{-\mathrm{NM}}=3$ translates into $n^\prime_+(2r^\prime -4q^\prime +1)=3-(2r^\prime +1)^2+8{q^\prime}^2$, which cannot be satisfied if $n^\prime_+$ is even \cite{foot2}.
When $n^\prime_+$ is odd instead, defining $k_0^\prime :=2k^\prime_O$ ($k^\prime_O\in \mathbb{Z}$), the condition 
$f_{-\mathrm{NM}}=3$ becomes 
$(p^\prime+k^\prime_O)(2r^\prime -4q^\prime +1)= 2q^\prime(q^\prime +1)-r^\prime(r^\prime +2)$.
Then we can rewrite $(B^\prime_0T)^2=-(16\pi^2/3)\{[2(p^\prime+k^\prime_O)+q^\prime +1]^2-3/8\}
({q^\prime}^2-3/8)$, and imposing the condition $(B_0T)^2\geq 0$ finally gives $q^\prime=0$ or $q^\prime=-[1+2(p^\prime +k^\prime_O)]$ and $(2r^\prime +1)[4(p^\prime+k^\prime_O) +2r^\prime +3]=3$.
The latter can be satisfied either if $p^\prime +k^\prime_O = 0$, together with $r^\prime =0, -2$, or if 
$p^\prime +k^\prime_O = -1$, together with $r^\prime =\pm 1$.

In conclusion, the time-optimal duration necessary to realize a $\mathrm{CNOT}(1, 3)$ gate with the Hamiltonian
\Ref{ising2} is given again by \Ref{topt}, while it is easy to check that the time-optimal magnetic fields $|B^\prime_0|, |B^\prime_z|$, the precession frequency  $\Omega$ and the ratio $J/|\omega|$  between the coupling $J$ and the energy parameter $\omega$ are also still expressed by, respectively, eqs. \Ref{b0z1}, \Ref{b0z2} \Ref{Omegaopt} and \Ref{jomega}. 


\section{Discussion}

We have presented the exact and analytical solution for the time-optimal realization of two entangling gates,
the $U^s_{13}$ and the $\mathrm{CNOT}(1, 3)$, between two indirectly coupled qubits, labelled 1 and 3, in a 3-qubit linear spin chain subject to, respectively, an Ising type interaction or a slightly modified Ising Hamiltonian, where in both cases a local magnetic field  can be applied on the intermediary qubit 2 and the constraint of a finite available energy is imposed.
No constraints ensuring instantaneous local unitary operations are imposed.
In particular, we showed that the $U^s_{13}$ gate can be optimally realized via an Ising Hamiltonian of the same form as that discussed in \cite{khaneja2} and that the time required is shorter than that found for the particular decomposition of the unitary evolution considered in \cite{khaneja2}.
We then presented the analytical solution for the time-optimal realization of the $\mathrm{CNOT}(1,3)$ via the slightly 
modified Ising Hamiltonian \Ref{ising2},  which again is shown to require the same time duration as the $U^s_{13}$ gate.

Of course the orbits generated by the Ising Hamiltonian \Ref{ising} cannot directly reach the target $\mathrm{CNOT}(1,3)$, which can instead be reached via the modified Ising Hamiltonian \Ref{ising2}.
According to the standard paradigm of time-optimal quantum computing (see, e.g., \cite{khaneja}), where one-qubit unitary operations are assumed to have zero time cost, one might infer that  the $\mathrm{CNOT}(1,3)$ can be still generated via the Ising Hamiltonian \Ref{ising} provided, for example, that a series of additional local operations acting on qubits 1, 2 and 3 are applied in the sequence given by formula (13) of \cite{khaneja}.
Somewhat surprisingly, this would appear to take the same time \Ref{topt}, which is shorter than that found in \cite{khaneja2}.
However, the reason of the better performance of our quantum evolution stems from the fact that the orbit described in \cite{khaneja2} is based on a special decomposition of the unitary operator as a product of unitary factors, each of which is indeed time-optimal, while the whole product would be time-optimal only if these single factors commuted which each other, which is not true in general.

On the other hand, in the more physical and general QB framework, where the time cost of local unitaries  is not negligible in principle, one can instead think of the new Hamiltonian \Ref{ising2} as generated by a change of basis of the Hilbert space for the three-qubit system according to $\{\ket{0}, \ket{1}\}\otimes \{\ket{0}, \ket{1}\}\otimes \{\ket{0}, \ket{1}\}\rightarrow  \{\ket{0}, \ket{1}\}\otimes \{\ket{0}, \ket{1}\}\otimes \{\ket{+}, \ket{-}\}$, as explained in the previous Section.
This would relate the Ising Hamiltonians \Ref{ising} and \Ref{ising2} via eq. \Ref{rotatedh}.
In other words, one can now see that the time-optimal realization of the gate $\mathrm{CNOT}(1,3)$
under the modified Hamiltonian \Ref{ising2} and where both the control and the target qubits belong to the standard computational basis $\{\ket{0}, \ket{1}\}$  is, in fact, equivalent to the time-optimal realization of the gate $\mathrm{CNOT}^{\pm}(1,3)$ under the Ising Hamiltonian \Ref{ising} where the control qubits belong to the basis $\{\ket{0}, \ket{1}\}$ while the target qubits belong to the basis  $\{\ket{+}, \ket{-}\}$.

We should note that  our problem is slightly more restricted than that of \cite{khaneja2}, in the sense that we also impose the finite energy condition eq. \Ref{eq-normH}.
In fact, the normalization condition \Ref{eq-normH} constrains not only the amplitude $|\vec{B}|$ of the local magnetic field in the Hamiltonian \Ref{ising} (or \Ref{ising2}), via eqs. \Ref{b0z1} and \Ref{b0z2}, but also the allowed values for the Ising coupling constant $J$, via eq. \Ref{jomega}.
As a consequence, only a (discrete) set of possible values for the coupling constant $J$ are actually allowed, and the optimal value of $J$ is related to the energy available $\omega$ through eq. \Ref{jomega}.
In particular, one can see that, in the case when the energy $\omega$ available increases (for a fixed integer $m$) the coupling $J$ grows and, therefore,
via eq. \Ref{topt}, the optimal time required to generate the gates decreases, while the strengths of the optimal local magnetic  field, via eqs. \Ref{b0z1} and \Ref{b0z2}, and also the precession frequency, via eq. \Ref{Omegaopt}, grow (which is what is physically expected).
Alternatively one can use eq. \Ref{jomega} to express the optimal values of $|B_0|$ and $|B_z|$ (eqs. \Ref{b0z1}  and \Ref{b0z2}) as a function of the total energy available $\omega$, and then to evaluate the ratio between the amplitude of the local magnetic field controlling the time-optimal evolution and the value of the Ising coupling constant $J$ available in the experiment, obtaining 
\begin{eqnarray}
\frac{|\vec{B}|}{J}=\frac{\pi}{2\sqrt{3}}\sqrt{5+ 8\left (m-\sqrt{\frac{3}{8}}\right )^2}.
\label{bj}
\end{eqnarray}
The ratio \Ref{bj} is a monotone increasing function of the integer $m$, and it tells us that the amplitude of the driving magnetic field  cannot be smaller than $|\vec{B}|_{\mathrm{MIN}} = 2\pi(1-\sqrt{3/8})^{1/2}/(\sqrt{3} J)$ (for $m=1$).
Furthermore, although the fact that the allowed values of $J$ belong to a discrete set (determined by the values of the integer $m$ via eq. \Ref{jomega}) might appear as a limitation of our model, we point out
that this feature is just the reflection of our simplifying technical choice (made just for ease of presentation) of 
considering only equality constraints \Ref{constraints}.
We conjecture that this feature should be mitigated by extending the QB methods to the more general and physical ansatz  where, e.g., the total energy available is only bounded from above (i.e., substituting the equality constraint \Ref{eq-normH} with an inequality constraint, and using the proper variational techniques, see, e.g., \cite{bertsekas}). 

Furthermore, we would like to comment on the relation of our results with those of \cite{complexity}.
In particular, in \cite{complexity} the one-qubit components of the time optimal Hamiltonian were found to be constant,
which is not the case in the present work. However, in \cite{complexity} there is no constraint for the one-qubit or the two-qubit components of the Hamiltonian, except for the normalization, which results in the absence of such components in $F'$ (eq. \Ref{eq-F}) and thus yields the constancy of the one-qubit components.
Here we have the constraints \Ref{cf1} \Ref{cf2} on the one-qubit and the two-qubit components of the Hamiltonian and therefore the result changes.

Our work was in part motivated by \cite{khaneja2}, and therefore we specialized our analysis to the case of the Hamiltonians \Ref{ising} and \Ref{ising2}, where the local operations are in principle allowed only on the intermediary qubit. 
However, there is no particular reason why one should limit to such a case, and one may think instead of more general situations in which also local interactions for (one of) the boundary qubits are available. 
Similarly, one could extend the analysis to the case of different interaction couplings between the indirectly coupled qubits (i.e. $J_{12}\not = J_{23}$, see \cite{yuan}) or for different topological couplings among the qubits (see, e.g., \cite{schulte} and \cite{fisher}).
Similarly, a simple and straightforward generalization of our results would be to include the study of how to time-optimally generate quantum gates other than the $\mathrm{CNOT}$ (e.g., the $\mathrm{SWAP}(1, 3)$, the $\sqrt{\mathrm{SWAP}(1, 3)}$
etc...). 
A further natural extension of our research will be also to consider our QB methods in the study of linear (Ising) chains of $n$ qubits (see \cite{yuan2}), e.g. in the search for time-efficient ways of creating and propagating spin order and coherences along the chains, or the transfer speed of single spin excitations via external magnetic fields along Heisenberg spin chains (see, e.g., \cite{murphy} and references therein).

\vspace{1cm}

\section*{ACKNOWLEDGEMENTS}
This research was partially supported by the MEXT of Japan, 
 under grant No. 09640341 (A.H. and T.K.).

\bibliographystyle{alpha}

\begin{thebibliography}{1}

\bibitem{brif} C. Brif., R. Chakrabarti and H. Rabitz, {\it New J. Phys.} {\bf 12}, 075008 (2010).


\bibitem{khaneja} N. Khaneja and S.J. Glaser, {\it Chem. Phys.} {\bf 267}, 11 (2001);
N. Khaneja, R. Brockett and S.J. Glaser, {\it Phys. Rev.} {\bf A63}, 032308 (2001).
\bibitem{dirr} G. Dirr, U. Helme, K. Hupr, M. Kleinsteuber and Y. Liu, {\it J. Glob. Opt.} {\bf 35}, 443 (2006).

\bibitem{vidal} G. Vidal, K. Hammerer and J.I. Cirac, {\it Phys. Rev. Lett.} {\bf 88}, 237902 (2002); id., {\it Phys. Rev.} {\bf A66}, 062321 (2002).
\bibitem{masanes} L. Masanes, G. Vidal and J.I. Latorre, {\it Quant. Inf. Comp.} {\bf 2}, 285 (2002).
\bibitem{zhang} J. Zhang, J. Vala, S. Sastry and K.B. Whaley, {\it Phys. Rev.} {\bf A67}, 042313 (2003).

\bibitem{kgb} N. Khaneja, S.J. Glaser and R. Brockett, {\it Phys. Rev.} {\bf A65}, 032301 (2002).
\bibitem{reiss} T.O. Reiss, N. Khaneja and S.J. Glaser, {\it J. Mag. Res.} {\bf 165}, 95 (2003).
\bibitem{yuan3} H. Yuan and N. Khaneja, {\it Phys. Rev.} {\bf A72}, 040301R (2005).

\bibitem{boscain} U. Boscain and P. Mason, {\it J. Math. Phys.} {\bf 47}, 062101 (2006);
U. Boscain, G. Charlot, J.P. Gauthier, S. Guerin and H.R. Jauslin, {\it J. Math. Phys.} {\bf 43}, 2107 (2002);
U. Boscain, I. Nikolaev and B. Piccoli, {\it J. Math. Sci.} {\bf 135}, 3109 (2006); 
U. Boscain and Y. Chitour, {\it SIAM J. Cont. Optim.} {\bf 44}, 111 (2005).
\bibitem{boscain2}
U. Boscain, T. Chambrion and G. Charlot, {\it Discr. Cont. Dyn. Sys} {\bf B5}, 957 (2005); 
U. Boscain, T. Chambrion and J.P. Gauthier, {\it J. Dyn. Contr. Sys.} {\bf 8}, 547 (2002).

\bibitem{sugny} D. Sugny, C. Kontz and H.R. Jauslin, {\it Phys. Rev.} {\bf A76}, 023419 (2007).

\bibitem{fisher} R. Fisher, H. Yuan, A. Sp\"orl and S. Glaser, {\it Phys. Rev.} {\bf A79}, 042304 (2009).

\bibitem{zeier} R. Zeier, H. Yuan and N. Khaneja, {\it Phys. Rev.} {\bf A77}, 032332 (2008).

\bibitem{childs} A. Childs, H. Haselgrove and M. Nielsen, {\it Phys. Rev.} {\bf A68}, 052311 (2003).

\bibitem{zeier2} R. Zeier, M. Grassl and T. Beth, {\it Phys. Rev.} {\bf A70}, 032319 (2004).
\bibitem{schulte}T. Schulte-Herbr\"uggen, A. Sp\"orl, N. Khaneja and S.J. Glaser, {\it Phys. Rev.} {\bf A72}, 042331 (2005).
\bibitem{nielsen3} M.A. Nielsen, {\it Quant. Inf. Comput.} {\bf 6}, 213 (2006).

\bibitem{nielsen1} M.A. Nielsen, M. Dowling, M. Gu and A. Doherty, {\it Science} {\bf 311}, 1133 (2006); id., {\it Phys. Rev.} {\bf A73}, 062323 (2006).
\bibitem{dowling} M. Dowling and M. A. Nielsen, {\it Quant. Inf. Comp.} {\bf 8}, 0861 (2008).
\bibitem{gu} M. Gu, A. Doherty and M.A. Nielsen, {\it Phys. Rev.} {\bf A78}, 032327 (2008).

\bibitem{tanimura} S. Tanimura, M. Nakahara and D. Hayashi, {\it J. Math. Phys.} {\bf 46}, 022101 (2005).


\bibitem{chuangnielsen} M.A. Nielsen and I.L. Chuang, {\it Quantum Computation and Quantum Information} (Cambridge University Press, Cambridge, 2000).

\bibitem{grape} N. Khaneja, T.O. Reiss, C. Kehlet, T. Schulte-Herbr\"uggen and S.J. Glaser {\it J. Mag. Res.} {\bf 172}, 296 (2005).
\bibitem{dominy} J. Dominy and H. Rabitz, {\it J.Phys.} {\bf A41}, 205305 (2008).

\bibitem{greek} From the Greek  ``$\beta\rho\alpha\chi\iota\sigma\tau o\zeta$", i.e. fast, and ``$\chi\rho o\nu o\zeta$", i.e time. 


\bibitem{pure} A. Carlini, A. Hosoya, T. Koike and Y. Okudaira, {\it Phys. Rev. Lett.} {\bf 96}, 060503 (2006).
\bibitem{mixed} A. Carlini, A. Hosoya, T. Koike and Y. Okudaira,   {\it J. Phys.}   {\bf A41}, 045303 (2008).
\bibitem{unitary} A. Carlini, A. Hosoya, T. Koike and Y. Okudaira,  {\it Phys. Rev.}  {\bf A75}, 042308 (2007).
\bibitem{fidelity} A. Carlini, A. Hosoya, T. Koike and Y. Okudaira,  {\it poster presented at AQIS 2007} (Kyoto, Japan).
\bibitem{complexity} T. Koike and Y. Okudaira, arXiv:0910.5587.
\bibitem{brodyhook} D. Brody and D. Hook, {\it J. Phys.} {\bf A39}, L167 (2006).

\bibitem{borras2} A. Borras, C. Zander, A.R. Plastino, M. Casas and A. Plastino, {\it Europhys. Lett.} {\bf 81}, 300007 (2008).
\bibitem{borras} A. Borras, A.P. Majtey and M. Casas, {\it Phys. Rev.} {\bf A78}, 022328 (2008).
\bibitem{zhao} B.K. Zhao, F.G. Deng, F.S. Zhang and H.Y. Zhou, {\it Phys. Rev.} {\bf A80}, 052106 (2009).

\bibitem{bbjm} C.M. Bender, D.C. Brody, H.F. Jones and B.K. Meister, {\it Phys. Rev. Lett.} {98}, 040403 (2007).
\bibitem{mostafa} A. Mostafazadeh, {\it Phys. Rev.} {\bf A79}, 014101 (2009).; id., {\it Phys. Rev. Lett.} {\bf 99}, 130502 (2007).
\bibitem{assis} P.E.G. Assis and A. Fring {\it J. Phys.} {\bf A41}, 244002 (2008).
\bibitem{fryd} A. Frydryszak and V. Tkachuk, {\it Phys. Rev.} {\bf A77}, 014103 (2008).
\bibitem{nesterov} A.I. Nesterov, {\it SIGMA} {\bf 5}, 069 (2009).
\bibitem{benbrod} C.M. Bender and D.C. Brody, {\it Time in Quantum Mechanics II} (Springer, Berlin, 2010).


\bibitem{khaneja2} N. Khaneja, B. Heitmann, A. Sp\" orl, H. Yuan, T. Schulte-Herbr\" uggen, and S.J. Glaser,  {\it Phys. Rev.}  {\bf A75}, 012322 (2007).
\bibitem{yuan} H. Yuan, R. Zeier and N. Khaneja, {\it Phys. Rev.} {\bf A77}, 032340 (2008).
\bibitem{yuan2} H. Yuan, S.J. Glaser and N. Khaneja, {\it Phys. Rev.} {\bf A76}, 012316 (2007).

\bibitem{kane} B.E. Kane, {\it Nature} {\bf 393}, 133 (1998); F. Yamaguchi, Y. Yamamoto,  {\it Appl. Phys. A: Mater. Sci. Process.}  {\bf 68}, 1 (1999).
\bibitem{mehring} M. Mehring, J. Mende and W. Scherer,  {\it Phys. Rev. Lett.}  {\bf 90}, 153001 (2003).
\bibitem{ernst} R.R. Ernst, G. Bodenhausen and A. Wokaun,  {\it Principles of Nuclear Magnetic Resonance in One and Two Dimensions} (Clarendon, Oxford, 1987).

\bibitem{invariance} The action $S$ is invariant under time-reparameterizations $\tau \rightarrow f(\tau)$.

\bibitem{otherway}  By choosing the Hamiltonian \Ref{ising} we are formally adopting the couplings $J_{12}(t)$ and $J_{23}(t)$ as dynamical variables in the action \Ref{eq-action}-\Ref{eq-LC}, and we make them 
become constant on shell via the imposition of the constraints \Ref{cf1}-\Ref{cf2}.
On the other hand, one might have chosen to start with $J_{12} =J_{23} := J =\mathrm{const}$, i.e. with the
coupling $J$ as a given constant parameter in the action \Ref{eq-action}-\Ref{eq-LC} from the beginning (cf. \cite{khaneja2}) and without the need of imposing the constraints \Ref{cf1}-\Ref{cf2}.
However, a simple calculation shows that both variational methods lead to the same result \Ref{bopt} for the time-optimal magnetic field ${\vec{B}}_{OPT}(t)$ (with $B_0$, $B_z$ and $\Omega$ given, respectively, by \Ref{b0z1}, \Ref{b0z2} and  \Ref{Omegaopt}) and to the same time-optimal duration \Ref{topt}.
In other words, our variational principle is essentially the same as that used in \cite{khaneja2}, with the caveat that we further impose the finite energy constraint \Ref{eq-normH}.   

\bibitem{b0} We exclude the trivial case of $B_0=0$.

\bibitem{us13} In fact, the condition $f_{-\mathrm{MIN}}=1$ is equivalent to $n_+(2r+1-4q)=2[2q(q -1)-r(r-1)]+1$, which cannot be satisfied if $n_+$ is even.

\bibitem{foot1} As  for $n_+=2(p+k_O)+1$ odd $f_{-\mathrm{NMM}}=3$ would imply that $2[(p+k_O)(2r-4q+1)+r^2-2q^2]=1$, which clearly is impossible.

\bibitem{foot2} Since, for $k^\prime := 2k^\prime_E$, one would have $2[(p^\prime +k^\prime_E)(2r^\prime -4q^\prime +1)+r^\prime(r^\prime +1) -2{q^\prime}^2]=1$, which is not possible.

\bibitem{bertsekas}  D.P. Bertsekas, {\it Nonlinear Programming} (Athena Scientific, Nashua, 1999).

\bibitem{murphy} M. Murphy, S. Montangero, V. Giovannetti and T. Calarco, {\it Phys. Rev.} {\bf A82}, 022318 (2010).




\end{thebibliography}

\end{document}